\documentclass[11pt]{article}
\usepackage{graphicx}

\newcommand{\BABARPubYear}    {01}

\newcommand{\BABARProcNumber} {57}
\newcommand{\SLACPubNumber} {8996}

\def\Journal#1#2#3#4{{#1} {\bf #2}, #3 (#4)}

\def\JHEP{\em J. High Energy Phys.}
\def\EPJC{{\em Eur. Phys. Jour.} C}

\def\NIM{\em Nucl. Instrum. Methods}

\def\PLB{{\em Phys. Lett.}  B}
\def\PRL{\em Phys. Rev. Lett.}
\def\PRD{{\em Phys. Rev.} D}

\def\BABAR{{\sc BaBar}}

\def\ra{\rightarrow}

\def\be{\begin{equation}}
\def\ee{\end{equation}}
\def\bea{\begin{eqnarray}}
\def\eea{\end{eqnarray}}

\input pubboard/babarsym

\long\def\inst#1{\par\nobreak\kern 4pt\nobreak
    {\it #1}\par\vskip 10pt plus 3pt minus 3pt}

\begin{document}
{\pagestyle{empty}

\begin{flushright}
SLAC-PUB-\SLACPubNumber \\
\babar-PROC-\BABARPubYear/\BABARProcNumber \\
October, 2001 \\
\end{flushright}

\par\vskip 4cm

\begin{center}
\Large \bf
Study of Hadronic and Rare B Decays in BABAR	
\end{center}
\bigskip

\begin{center}
\large 
L. Lista\\
INFN Sezione di Napoli \\
Complesso Universitario di Monte Sant'Angelo \\
 via Cintia, 80126 Napoli, Italy \\
E-mail: {\tt luca.lista@na.infn.it}\\ 
(for the \lbabar\ Collaboration)
\end{center}
\bigskip \bigskip

\begin{center}
\large \bf Abstract
\end{center}
We present results from \BABAR\ experiment for the measurement of
inclusive and exclusive branching fractions of B mesons into final
states containing $J/\psi$, $\psi(2S)$ and $\chi_c$. The contributions of CP even
and odd amplitudes in the decay $B^0 \rightarrow J/\psi K^{*0}$ are determined from an
angular analysis. We report the measurements of the branching ratios $B^0
\rightarrow D^{*+}D^{*-}$ and $D^{*+}D^{*-}K^0_S$, and the study of exclusive two-body and
quasi-two-body charmless decays. The branching fraction of the decay 
$B^0 \rightarrow K^{*0} \gamma$ has been determined and the corresponding CP asymmetry has
been measured.

\vfill
\begin{center}
Contributed to the Proceedings of the International
Conference on Flavor Physics (ICFP 2001)\\
31 May - 6 Jun 2001, Zhang-Jia-Jie City, Hunan, China
\end{center}

\vspace{1.0cm}
\begin{center}
{\em Stanford Linear Accelerator Center, Stanford University, 
Stanford, CA 94309} \\ \vspace{0.1cm}\hrule\vspace{0.1cm}
Work supported in part by Department of Energy contract DE-AC03-76SF00515.
\end{center}

\section{PEP-II and \BABAR\ detector}

The PEP-II asymmetric B-factory collides electrons with 9~GeV energy and
positrons with 3.1~GeV energy. The center-of-mass energy corresponds to 
the $\Upsilon(4S)$ resonance, which is produced with a 
Lorentz boost of $\beta\gamma = 0.56$ in the laboratory frame.
The peak luminosity reached in the running period from November 1999 to October 2000
was $3.3\times 10^{33}$~cm$^{-2}$s${-1}$. The integrated luminosity in that period
is approximately $21$~fb$^{-1}$. An additional $4$~fb$^{-1}$ have been collected
below the resonance, in order to study the non-resonant $e^+e^-\ra q\bar q$ background.

The \BABAR\ detector~\cite{babar-nim} consists of a five-layer double-sided Silicon
 Vertex Tracker (SVT), a Drift Chamber (DCH) with 40 layer, a Cherenkov radiation 
detector (DIRC: Detector of Internally Reflected Cherenkov light)~\cite{dirc}, 
a finely-segmented CsI(Tl) Electromagnetic Calorimeter (EMC) 
and a muon and neutral hadron detector (IFR: Instrumented Flux
Return) consisting of Resistive Plate Chambers (RPC)
incorporated in the flux return of the 1.5~T magnetic field produced by a
solenoidal superconducting magnet.

\section{B counting}

The determination of the branching fractions of B decays
requires a correct estimate of the number of produced
$\Upsilon(4S)\ra BB$ events. 
The number of $BB$ events, $N_{BB}$, is estimated from a
selection of multihadronic decays, after subtracting the expected
number of events from continuum, which is
determined from PEP-II runs taken at a center-of-mass energy below
the $\Upsilon(4S)$ resonance, scaled according to the number of selected dimuon
events recorded in on- and off-resonance data. The number of $BB$ events 
is:
\begin{equation}
N_{BB} = N_{MH} - N_{MH}^{off} \frac{N_{\mu\mu}}{N_{\mu\mu}^{off}}\kappa,
\end{equation}
where $N_{MH}$ is the number of multihadronic events taken in on-resonance runs,
$N_{MH}^{off}$ is the number of multihadron events 
selected off-resonance, $N_{\mu\mu}$ and $N_{\mu\mu}^{off}$ are the 
number of selected dimuon events in on- and off-resonance data respectively.
The term $\kappa=0.9962\pm0.0027$ corrects for the different
efficiencies at different center-of-mass energies
for dimuon and multihadronic events.
The number of produced $\Upsilon(4S)\ra BB$ decays for the
analyses that are presented in this paper is $(22.6\pm 0.4)\times 10^{6}$ events.

\section{Charmonium decays}

Many $B$ decays containing charmonium final states 
are important for $CP$ violation studies and the
determination of $\sin 2\beta$. Moreover, 
the precise determination of many inclusive and exclusive
branching fractions can be used for a test
of theoretical predictions based on the factorization 
hypothesis.

$J/\psi$, $\psi(2S)$ and $\chi_c$ candidates are 
reconstructed in final states containing leptons. 
$J/\psi$ candidates are reconstructed in decays 
to $\ell^+\ell^-$ ($\ell$ = $e$ or $\mu$), 
$\psi(2S)$ candidates are reconstructed both in 
$\ell^+\ell^-$ and $J/\psi\pi^+\pi^-$ decay modes, and $\chi_c$ 
candidates are 
reconstructed in the decay mode $J/\psi\gamma$. 
Details of the performance of \BABAR\ for muon 
and electron identification and photon energy 
measurement can be found in the Ref.~\cite{babar-nim}.

In the case of final states containing electrons,
a bremsstrahlung recovery algorithm is applied:
final state radiation photons that might have been
radiated from electrons coming from charmonium 
decays are identified and considered in 
the calculation of the final state kinematics.

\subsection{Inclusive Charmonium decays}

The inclusive branching ratios of $B$ decays to charmonium final states 
are obtained from 
the fit to the number of events under the peaks of charmonium resonances, 
correct for particle identification and reconstruction efficiencies (estimated
from control samples extracted from real data), normalized to
the number of produced $BB$ events estimated with the
$B$-counting procedure described in the previous section.

Inclusive branching fractions for direct decays of
$B\ra J/\psi X$ and $B\ra \chi_{c1} X$ are 
obtained subtracting the contribution expected from
the decay $B\ra \psi(2S) X$ followed by a decay of the $\psi(2S)$
in $J/\psi$ and $\chi_{c1}$, respectively. The
branching fractions for $\psi(2S)$ to decay to $J/\psi$ and $\chi_{c1}$
are assumed to be the PDG values.
We determine the inclusive branching fractions shown in 
Table~\ref{tab:charmincl}. The $\chi_c$ analysis shows evident for
$\chi_{c1}$ resonance, with a shoulder that could be interpreted as 
$\chi_{c2}$. The branching fraction for $B\ra \chi_{c2}X$ is determined to be
$0.137\pm 0.058\pm 0.012$, but given 
the low statistical significance, we prefer to quote
the upper limit $Br(B\ra \chi_{c2}X) < 0.21\times 10^{-2} (90\%)$~C.L.\, .

\begin{table}[htbp]
\caption{Measured inclusive branching fraction for $B$ to decay to charmonium
final states. The last two measurement refer to inclusive branching fractions for direct decay.
\label{tab:charmincl}}
\vspace{0.2cm}
\begin{center}
\footnotesize
\begin{tabular}{|c|c|}
\hline
\raisebox{0pt}[13pt][7pt]{Decay Mode} & Branching Fraction\\
\hline
\raisebox{0pt}[13pt][7pt]{$B\ra J/\psi X$} & $(1.044\pm 0.013 \pm 0.028) \times 10^{-2}$\\
\raisebox{0pt}[13pt][7pt]{$B\ra \psi(2S) X$} & $(0.275\pm 0.020 \pm 0.029) \times 10^{-2}$\\
\raisebox{0pt}[13pt][7pt]{$B\ra \chi_{c1} X$} & $(0.378\pm 0.034 \pm 0.026) \times 10^{-2}$\\
\raisebox{0pt}[13pt][7pt]{$B\ra \chi_{c2} X$} & $< 0.21 \times 10^{-2}, 90\%$ C.L.\\
\hline
\raisebox{0pt}[13pt][7pt]{$B\ra J/\psi X (dir.)$} & $(0.789\pm 0.010 \pm 0.034) \times 10^{-2}$\\
\raisebox{0pt}[13pt][7pt]{$B\ra \chi_{c1} X (dir.)$} & $(0.353\pm 0.034 \pm 0.024) \times 10^{-2}$\\
\hline
\end{tabular}
\end{center}
\end{table}

As a byproduct of this analysis, we determine
the branching fractions for $\psi(2S)\ra e^+e^-$ and
$\psi(2S)\ra\mu^+\mu^-$ from the ratio of branching 
fractions for $B\ra\psi(2S)X\ra e^+e^-X$ and 
$B\ra\psi(2S)X\ra \mu^+\mu^-X$ to $B\ra\psi(2S)X\ra J/\psi\pi^+\pi^-X$, 
assuming the PDG value for the branching
ratio $\psi(2S)\ra J/\psi\pi^+\pi^-$.

The results are
\begin{eqnarray*}
 {\cal B}(\psi(2S)\ra e^+e^-) &= &(8.1 \pm 0.9 \pm 0.9)\times10^{-3}, \\
 {\cal B}(\psi(2S)\ra \mu^+\mu^-) &= &(7.0 \pm 0.8 \pm 0.9)\times10^{-3}.
\end{eqnarray*}
The $\psi(2S)\ra \mu^+\mu^-$ measurement significantly
improves the precision of the current world average.

\subsection{Charmonium production in continuum}
 
The sample of $J/\psi$ candidates shows evidence for
events in which the momentum of the
$J/\psi$ is larger than 2~GeV in the $\Upsilon(4S)$ rest frame,
which is kinematically forbidden for $B$ decays.
The observed number of events is consistent with observed number of
$J/\psi$ candidates in the data taken below the $\Upsilon(4S)$ resonance
energy~\cite{jpsicont}.
%
Those events are interpreted as $J/\psi$ 
production in continuum.

We determine a production cross section of
$\sigma_{e^+e^-\ra J/\psi X} = (2.52 \pm 0.21 \pm 0.21)$~pb and the direct
production from $\Upsilon(4S)$ is excluded: ${\cal B}(\Upsilon(4S)\ra J/\psi X) 
< 4.3\times 10^{-4}, 90\%$~C.L.\,.
The measured cross section together with the observed 
angular distribution of the $J/\psi$~\cite{jpsicont},
favours the non-relativistic QCD~\cite{nrqcd} 
models over color-singlet models~\cite{colsing}.

The (small) contribution from $J/\psi$ production in continuum events has been 
subtracted from the estimates of inclusive charmonium branching
fractions reported in Table~\ref{tab:charmincl}.

\subsection{Exclusive Charmonium decays}

The kinematic selection of exclusive $B$ decays is based on two main
variables that provide an effective background discrimination.
\begin{itemize}
\item{} The energy substituted mass:
\[
     m_{ES} = \sqrt{{s\over 4}-p^2_{B,cm}}\, ,
\]
where $s$ is the PEP-II center-of-mass energy squared, and $p_{B,cm}$ is
the measured $B$ momentum in the $\Upsilon(4S)$ rest frame.
In the calculation of $m_{ES}$ there is no assumption on the mass of the particles
in the $B$ final state,
so  $m_{ES}$ doesn't depend on particle identification.

\item{} Energy difference:
\[
     \Delta E = E_{B, cm} - {\sqrt{s} \over 2}
\]

where $E_{B,cm}$ is the measured $B$ energy in the $\Upsilon(4S)$ rest frame.
The $\Delta E$ distribution peaks around 0 for correctly reconstructed $B$ decays.
\end{itemize}

In the decay $B^0\ra J/\psi K^0_L$ the $K^0_L$ is reconstructed
as a cluster in the IFR and/or in the EMC, so its direction is measured,
but no energy determination is possible. 
Therefore, only the variable $ \Delta E$ is determined, assuming the
mass constraint $m_{ES} = m_B$ to determine the momentum of the $K_L$.

We measure~\cite{charmexcl} the branching fractions summarised in Table~\ref{tab:charmexcl}.
We report the first observation of $B^0\ra \chi_{c1}K^{0*}$.

\begin{table}[!htb]
\caption{Branching fraction for $B$ decays to exclusive charmonium
final states.
\label{tab:charmexcl}}
\vspace{0.2cm}
\begin{center}
\footnotesize
\begin{tabular}{|c|c|}
\hline
\raisebox{0pt}[13pt][7pt]{Decay Mode} & Branching Fraction\\
\hline
\raisebox{0pt}[13pt][7pt]{$B^0 \ra J/\psi \pi^0$}&$(0.20 \pm 0.06 \pm 0.02)\times 10^{-4} $\\
\raisebox{0pt}[13pt][7pt]{$B^0 \ra J/\psi K^{*0}$}&$(12.4 \pm 0.5 \pm 0.9)\times 10^{-4}$\\
\raisebox{0pt}[13pt][7pt]{$B^+ \ra J/\psi K^{*+}$}&$(13.7 \pm 0.9 \pm 1.1)\times 10^{-4}$\\
\raisebox{0pt}[13pt][7pt]{$B^+ \ra  J/\psi K^+$}&$(10.1 \pm 0.3 \pm 0.5)\times 10^{-4}$\\
\raisebox{0pt}[13pt][7pt]{$B^0 \ra J/\psi K^0 (K_L)$}&$(6.8 \pm 0.8 \pm 0.8)\times 10^{-4}$\\
\raisebox{0pt}[13pt][7pt]{$B^0 \ra J/\psi K^0 (K_S\ra\pi^0\pi^0)$}&$(9.6 \pm 1.5 \pm 0.7)\times 10^{-4} $\\
\raisebox{0pt}[13pt][7pt]{$B^0 \ra J/\psi K^0 (K_S\ra\pi^+\pi^-)$}&$(8.5 \pm 0.5 \pm 0.6)\times 10^{-4}$\\
\raisebox{0pt}[13pt][7pt]{$B^0 \ra J/\psi K^0 (All)$}&$(8.3 \pm 0.4 \pm 0.5)\times 10^{-4}$\\
\raisebox{0pt}[13pt][7pt]{$B^0 \ra \chi_{c1} K^{*0}$}&$(4.8 \pm 1.4 \pm 0.9)\times 10^{-4}$\\
\raisebox{0pt}[13pt][7pt]{$B^0 \ra \chi_{c1} K^0$}&$(5.4 \pm 1.4 \pm 1.1)\times 10^{-4}$\\
\raisebox{0pt}[13pt][7pt]{$B^+ \ra \chi_{c1} K^+$}&$(7.5 \pm 0.8 \pm 0.8)\times 10^{-4}$\\
\raisebox{0pt}[13pt][7pt]{$B^+ \ra \psi(2S) K^+$}&$(6.3 \pm 0.5 \pm 0.8)\times 10^{-4}$\\
\raisebox{0pt}[13pt][7pt]{$B^0 \ra \psi(2S) K^0$}&$(6.8 \pm 1.0 \pm 1.1)\times 10^{-4}$\\
\hline
\end{tabular}
\end{center}
\end{table}

Using an unbinned maximun likelihood fit based on kinematic 
variables, we also determine~\cite{fabozzi} the ratio
of branching fractions for the decays $B^+\ra J/\psi \pi^+$
and $B^+\ra J/\psi K^+$ to be
\[
{{\cal B}(B^+\ra J/\psi \pi^+) \over {\cal B}(B^+\ra J/\psi K^+)} =(3.91\pm0.78 \pm0.19)\times 10^{-2}.
\]

\subsection{$J/\psi K^{*0}$ angular analysis}

The decay $B^0\ra J/\psi K^{*0}$
contains a mixture of $CP$-odd and
$CP$-even amplitudes. It is possible to 
interpret CP-violation asymmetries in this channel
by disentangling the two CP
components with an angular analysis
of the decay products.

The decay can be characterized by three amplitudes $A_\parallel$, 
$A_0$ and $A_\perp$ (defined in the Ref.~\cite{angular}), which correspond to 
the $CP$-odd P-wave ($A_\perp$) and $CP$-even mixtures of S and D waves
($A_0$ and $A_\parallel$).
The squared moduli of the three amplitudes and their relative phases
can be determined from a fit to the angular distribution of
the decay products.

\BABAR\ determinations from fits to the angular distributions are
\begin{eqnarray*}
|A_\perp|^2     & = & 0.160 \pm 0.032 \pm 0.014\, , \\
|A_\parallel|^2 & = & 0.243 \pm 0.034 \pm 0.017\, , \\
|A_0|^2         & = & 0.597 \pm 0.028 \pm 0.02\, , \\
\phi_\perp     & = & \arg \left({A_\perp \over A_0}\right) = -0.17 \pm 0.16 \pm 0.07\, , \\
\phi_\parallel & = & \arg \left({A_\parallel \over A_0}\right) = 2.50 \pm 0.20 \pm 0.08\, .
\end{eqnarray*}

In the factorization hypothesis the relative phases of
the three amplitudes should be $0$ or $\pi$.
The measurement of $\phi_\parallel$ shows a deviation
from $\pi$ of three standard deviations, giving indication of
the presence of final-state interactions.

The dilution factor for the measurement of
$\sin 2 \beta$ in this
channel is determined from the relative $CP$-odd contribution:
\[
 D = 1 - 2 |A_\perp|^2 = 0.68\pm 0.10.
\]
\section{Open Charm decays}

$B$ mesons decays to three-body final states through the chain $b\ra c\bar cs$ have
been studied by \BABAR\ . 
The current experimental results on two-body 
decays $B\ra D_sX$, $(cc)X$, $\Lambda_cX$, $\Xi_cX$
from ALEPH~\cite{aleph} and CLEO~\cite{cleo}
give a contribution to a $b\ra c\bar cs$ branching
fraction~\cite{bccs} lower than expected from the theoretical extrapolation
from the semileptonic branching ratio.
This may be an indication that the three-body decays could give a non-negligible 
contribution to the total $b\ra c\bar cs$ branching
fraction.
\BABAR\ results on $B\ra D^*D^{(*)}K$ modes are the following:
\begin{eqnarray*}
 {\cal B}(B^0\ra D^{*+}D^0K^+)     & = & (0.28\pm 0.07\pm 0.05)\times 10^{-2}\, , \\
 {\cal B}(B^0\ra D^{*+}D^{*0}K^+) & = & (0.68\pm 0.17\pm 0.17)\times 10^{-2}\, , \\
 {\cal B}(B^+\ra D^{*+}D^{*-}K^+) & = & (0.34\pm 0.16\pm 0.11)\times 10^{-2}\, .
\end{eqnarray*}
The last mode is the first experimental evidence of a 
color-suppressed mode other than $B\ra$(charmonium)$X$.

The decays $B\ra D^{*+}D^{*-}$ provide an interesting 
Cabibbo-suppressed $b\ra c\bar c d$ process that could 
provide, with enough events, a determination of $\sin 2\beta$
independent on the mode $J/\psi K^0$. The determination
of $\sin 2\beta$ would require particular care in the 
subtraction of penguin contributions, which could be non-negligible.
The theoretical estimate for the branching fraction of
this mode is
\[
  {\cal B}(B\ra D^{(*)}\bar{D}^{(*)})\approx \left(
   {f_{D^{*}}\over f_{D^{*}_s}} \right) \tan\theta_c 
 {\cal B}(B\ra D^{(*)}_s\bar{D}^{(*)})\approx 0.1\%\, .
\]
\BABAR\ measures
\[
  Br(B^0\ra D^{*+}D^{*-}) = (8.0 \pm 1.6\pm 1.2) \times 10^{-4}\, .
\]

\section{Two-body decays} 

In the Standard Model the study of charmless two-body $B$ decays 
with final states $\pi\pi$ and $K\pi$ 
allows us to determine, with sufficient events, through a time dependent CP
asymmetry study, the value of $\sin 2\alpha$.
Unlike the $\sin 2\beta$ modes, these decays
have a penguin contribution that is non-negligible
compared to the Cabibbo-suppressed tree decay,
and whose phase differs from the tree diagram phase.
The parameter $\sin 2\alpha$ can be extracted
frrom the raw CP-violating asymmetry
for $B^0\ra \pi^+\pi^-$ by applying 
a correction that takes into account the penguin pollution~\cite{alpha}.
This correction depends on 
all the other $B\ra \pi\pi$ amplitudes 
and, in particular, both $B^0\ra \pi^0\pi^0$ 
and $\bar B^0\ra \pi^0\pi^0$ need 
to be measured.

Particle identification with the DIRC is 
of fundamental importance to correctly
discriminate pions from kaon background.
Continuum events are rejected on the basis
of event shape variables that are combined
into a Fisher discriminant.
The branching fractions are determined from
an unbinned likelihood fit that
combines kinematical information with
the Cherenkov angle measured in the DIRC.
The probability distribution functions for the
Cherenkov angle have been determined 
from the control sample $D^{*+}\ra D^0\pi^+\ra (K^-\pi^+)\pi^+$
in real data, and for the kinematical variables
from the off-resonance data and the
($\Delta E$, $m_{ES}$) sidebands.
The results for all two-body modes are 
summarized in the Tables~\ref{tab:twobodyhh},
\ref{tab:twobodypi0h} and \ref{tab:twobodyk0h}.

\begin{table}[htbp]
\caption{$B^0\ra h^+h^-$ branching fractions.
\label{tab:twobodyhh}}
\vspace{0.2cm}
\begin{center}
\footnotesize
\begin{tabular}{|c|c|c|c|}
\hline
\raisebox{0pt}[13pt][7pt]Decay Mode & Branching Fraction &  Yield & Significance\\
\hline
\raisebox{0pt}[13pt][7pt]{$B^0\ra K^+\pi^-$} & 
 $(16.7\pm 1.6^{+1.2}_{-1.7})\times 10^{-6}$ & $169\pm 17 ^{+12}_{-17}$
&$15.8\sigma$ \\

\raisebox{0pt}[13pt][7pt]{$B^0\ra \pi^+\pi^-$} & 
 $(4.1\pm 1.0\pm 0.7)\times 10^{-6}$ & $41\pm 10 \pm 7$
&$4.7\sigma$ \\

\raisebox{0pt}[13pt][7pt]{$B^0\ra K^+K^-$} & 
 $ < 2.5\times 10^{-6}$ (90\%C.L.) & $8.2^{+7.8}_{-6.4}\pm 3.3$
&$1.3\sigma$ \\

\hline
\end{tabular}
\end{center}
\end{table}
\begin{table}[htbp]
\caption{$B^+\ra \pi^0h^+$ branching fractions.
\label{tab:twobodypi0h}}
\vspace{0.2cm}
\begin{center}
\footnotesize
\begin{tabular}{|c|c|c|c|}
\hline
\raisebox{0pt}[13pt][7pt]Decay Mode & Branching Fraction & Yield & Significance\\
\hline
\raisebox{0pt}[13pt][7pt]{$B^+\ra \pi^0\pi^+$} & 
 $(5.1^{+2.0}_{-1.8}\pm 0.8)\times 10^{-6}$ & $37^{+15}_{-13}$
&$3.4\sigma$ \\

\raisebox{0pt}[13pt][7pt]{$B^+\ra \pi^0K^+$} & 
 $ (10.8^{+2.1}_{-1.9}\,^{+1.0}_{-1.2})\times 10^{-6}$ & $75^{+14}_{-13}$
&$8.0\sigma$ \\

\hline
\end{tabular}
\end{center}
\end{table}
\begin{table}[htbp]
\caption{$B^+\ra K^0h^+$ and $B^0\ra K^0\pi^0$ branching fractions.
\label{tab:twobodyk0h}}
\vspace{0.2cm}
\begin{center}
\footnotesize
\begin{tabular}{|c|c|c|c|}
\hline
\raisebox{0pt}[13pt][7pt]Decay Mode & Branching Fraction &  Yield & Significance\\
\hline
\raisebox{0pt}[13pt][7pt]{$B^+\ra K^0\pi^+$} & 
 $(18.2^{+3.3}_{-3.0}\,^{+1.6}_{-2.0})\times 10^{-6}$ & $59^{+11}_{-10}$
&$9.8\sigma$ \\

\raisebox{0pt}[13pt][7pt]{$B^+\ra K^0K^+$} & 
 $<2.6\times 10^{-6}$ (90\% C.L.)& $0\,\, (<8,$ 90\%C.L.)
&$0\sigma$ \\

\raisebox{0pt}[13pt][7pt]{$B^0\ra K^0\pi^0$} & 
 $(17.9^{+6.8}_{-5.8}\,^{+1.1}_{-1.2})\times 10^{-6}$ & $17.99^{+6.8}_{-5.8}$
&$4.5\sigma$ \\

\hline
\end{tabular}
\end{center}
\end{table}

The decay channels for the $B^\pm$ and $B^0\ra K^+\pi^-$
are self-tagging, in the sense that the flavor of the $b$ quark
in the $B$ meson can be determined from the charge of 
the particles in the final state.
For those decays we have determined the direct CP asymmetries reported
in Table~\ref{tab:dircp}.

\begin{table}[htbp]
\caption{Direct CP asymmetries.
\label{tab:dircp}}
\vspace{0.2cm}
\begin{center}
\footnotesize
\begin{tabular}{|c|c|c|c|}
\hline
\raisebox{0pt}[13pt][7pt]Decay Mode & CP Asymmetry &  90\% C.L. interval\\
\hline
\raisebox{0pt}[13pt][7pt]{$B^0\ra K^+\pi^-$} & 
 $-0.19\pm0.10\pm0.03$ & $[-0.35,+0.03]$ \\

\raisebox{0pt}[13pt][7pt]{$B^+\ra K^+\pi^0$} & 
 $0.00\pm0.18\pm0.04$ & $[-0.30,+0.30]$ \\

\raisebox{0pt}[13pt][7pt]{$B^-\ra K^0\pi^-$} & 
 $-0.21\pm0.18\pm0.03$ & $[-0.51,+0.09]$ \\

\hline
\end{tabular}
\end{center}
\end{table}

The decays $B\ra \phi K$ and $B\ra \phi K^*$ are dominated 
by the gluonic penguin diagrams for $b\ra s\bar s s$, and 
provide potentially an independent determination
of $\sin 2\beta$ and could show direct CP violation effects.

The $\phi$ is reconstructed in the decay mode $\phi\ra K^+K^-$,
where both final state kaons are identified in the DIRC and 
must have an invariant mass within $\pm30$~MeV of the $\phi$ mass.
\BABAR\ results for $B\ra \phi K^{(*)}$ and $\phi \pi$ are reported 
in Table~\ref{tab:phik}.

\begin{table}[htbp]
\caption{$B^+\ra \phi K^{(*)}, \phi\pi$ branching fractions.
\label{tab:phik}}
\vspace{0.2cm}
\begin{center}
\footnotesize
\begin{tabular}{|c|c|c|c|}
\hline
\raisebox{0pt}[13pt][7pt]Decay Mode & Decay Fraction &  Yield & Significance\\
\hline
\raisebox{0pt}[13pt][7pt]{$B^+\ra \phi K^+$} & 
 $(7.7^{+1.6}_{-1.4}\pm 0.8)\times 10^{-6}$ & $31.4^{+6.7}_{-5.9}$
&$10.5\sigma$ \\

\raisebox{0pt}[13pt][7pt]{$B^+\ra \phi \pi^+$} & 
 $<1.4\times 10^{-6} (90\%C.L.)$ & $0.9^{+2.1}_{-0.9}$
&$0.6\sigma$ \\

\raisebox{0pt}[13pt][7pt]{$B^+\ra \phi K^{*+}$} & 
 $(9.7^{+4.2}_{-3.4}\pm 1.3)\times 10^{-6}$ & $-$
&$4.5\sigma$ \\

\hline

\raisebox{0pt}[13pt][7pt]{$B^+\ra \phi K^{*+}(K^+)$} & 
 $(12.8^{+7.7}_{-6.1}\pm 3.2)\times 10^{-6}$ & $7.1^{+4.3}_{-3.4}$
&$2.7\sigma$ \\

\raisebox{0pt}[13pt][7pt]{$B^+\ra \phi K^{*+}(K^0)$} & 
 $(8.0^{+5.0}_{-3.7}\pm 1.3)\times 10^{-6}$ & $4.4^{+2.7}_{-2.0}$
&$3.6\sigma$ \\
\hline

\raisebox{0pt}[13pt][7pt]{$B^0\ra \phi K^{*0}$} & 
 $(8.6^{+2.8}_{-2.4}\pm 1.1)\times 10^{-6}$ & $16.9^{+5.5}_{-4.7}$
&$6.6\sigma$ \\

\raisebox{0pt}[13pt][7pt]{$B^0\ra \phi K^0$} & 
 $(8.1^{+3.1}_{-2.5}\pm 0.8)\times 10^{-6}$ & $10.8^{+4.1}_{-3.3}$
&$6.4\sigma$ \\

\hline
\end{tabular}
\end{center}
\end{table}

\section{Quasi-two body decays} 

A number of quasi-two-body and three-body decay modes have been studied
by \BABAR. Of particular interest, a three-body Dalitz analysis 
of $B\ra \rho\pi$ decays allows a determination of $\sin 2\alpha$ 
and, at the same time, the strong phases of the tree and 
penguin diagrams~\cite{rhopi}.
\BABAR\ results for various modes are summarized in Table~\ref{tab:quasi}.

\begin{table}[htbp]
\caption{Quasi-two-body and three-body branching fractions.
\label{tab:quasi}}
\vspace{0.2cm}
\begin{center}
\footnotesize
\begin{tabular}{|c|c|c|c|}
\hline
\raisebox{0pt}[13pt][7pt]Decay Mode & Branching Fraction \\
\hline
\raisebox{0pt}[13pt][7pt]{$B^+\ra \omega h^+$} &  $<24\times10^{-6} (90\%$ C.L. \\
\raisebox{0pt}[13pt][7pt]{$B^0\ra \omega K^0$} &  $<14\times10^{-6} (90\%$ C.L. \\
\raisebox{0pt}[13pt][7pt]{$B^+\ra \eta^\prime K^+$} &  $(62\pm18\pm8)\times10^{-6}$ \\
\raisebox{0pt}[13pt][7pt]{$B^0\ra \eta^\prime K^0$} &  $<112\times10^{-6} (90\%C.L.)$ \\
\raisebox{0pt}[13pt][7pt]{$B^+\ra K^{*0}\pi^+$} &  $<28\times10^{-6} (90\%C.L.)$ \\
\raisebox{0pt}[13pt][7pt]{$B^+\ra \rho^0K^+$} &  $<39\times10^{-6} (90\%C.L.)$ \\
\raisebox{0pt}[13pt][7pt]{$B^+\ra \rho^0\pi^+$} &  $<39\times10^{-6} (90\%C.L.)$ \\
\raisebox{0pt}[13pt][7pt]{$B^+\ra K^+\pi^+\pi^-$} &  $<54\times10^{-6} (90\%C.L.)$ \\
\raisebox{0pt}[13pt][7pt]{$B^+\ra \pi^+\pi^+\pi^-$} &  $<22\times10^{-6} (90\%C.L.)$ \\
\raisebox{0pt}[13pt][7pt]{$B^0\ra \rho^\pm\pi^\mp$} &  $(49\pm 13^{+6}_{-5})\times 10^{-6}$ \\

\hline
\end{tabular}
\end{center}
\end{table}

\section{Radiative penguin decays}

The decay $B\ra K^*\gamma$ is dominated by the electromagnetic penguin diagram.
The main contribution comes from the diagram involving a $t$ quark in the loop,
so it is sensitive to the CKM elements $V_{ts}$ and $V_{td}$.
The Standard Model prediction for the direct CP asymmetry
in $B\ra K^*\gamma$ is below 1\%.

In the presence of new physics beyond the Standard Model, the electromagnetic 
penguin could be enhanced by charged Higgs or supersymmetric particles,
increasing the branching ratio.
The direct CP asymmetry could become measurable at \BABAR.

The selection of $B^0\ra K^{*0}\gamma$ events is based on identification of 
a photon in the calorimeter, not coming from the decay of a
$\pi^0$ or $\eta$, and  reconstruction of
$K^{*0}$ candidates decaying to $K^+\pi^-$. 
The rejection of the non-resonant $e^+e^-\ra q\bar q$
background is performed on the basis of the angle
formed by the photon and the thrust axis
of all particles in the event except those belonging to the 
$K^*\gamma$ candidate.
\BABAR\ determines the following branching ratio:
\[
  {\cal B}(B^0\ra K^{*0}\gamma) = (4.39\pm 0.41 \pm 0.27)\times 10^{-5}\, .
\]
The sample is composed of $72\pm 9$
 $B^0\ra K^{*0}\gamma \ra K^+\pi^-\gamma$ events and 
$67.2\pm 9.1$ $\bar B^0\ra \bar K^{*0}\gamma \ra K^+\pi^-\gamma$ 
events. The direct
CP asymmetry has been determined to be
\[
  A_{CP}(B^0\ra K^{*0}\gamma) = -0.035 \pm 0.094 \pm 0.022\, .
\]

\section{Conclusions}

\BABAR\ has recorded 22.6 million $BB$ events in the
period from November 1999 to October 2000.
Analyses hadronic $B$ decays and the study of rare processes in \BABAR\ have shown
first evidence of charmonium production in continuum and have 
provided precise determinations of
branching fractions of $B$ to exclusive and inclusive charmonium 
final states, including the first observation of the decay $B^0 \ra \chi_{c1}K^{*0}$.
\BABAR\ has observed the color-suppressed decay $B^+\ra D^{*+}D^{*-}K^+$,
and studied two-body charmless decays that will provide, with larger
data samples, a measurement of $\sin 2\alpha$.
The radiative penguin decay $B^0\ra K^{*0}\gamma$ has been studied and
a first measurement of the $CP$ asymmetry in that channel has 
been performed.

Many measurements are still dominated by statistical uncertainty, 
and more events will allow the reduction of the errors and access
to more rare channels.

\section*{Acknowledgments}
We are grateful for the 
extraordinary contributions of our PEP-II colleagues in
achieving the excellent luminosity and machine conditions
that have made this work possible.
The collaborating institutions wish to thank 
SLAC for its support and the kind hospitality extended to them. 
This work is supported by the
US Department of Energy
and National Science Foundation, the
Natural Sciences and Engineering Research Council (Canada),
Institute of High Energy Physics (China), the
Commissariat \`a l'Energie Atomique and
Institut National de Physique Nucl\'eaire et de Physique des Particules
(France), the
Bundesministerium f\"ur Bildung und Forschung
(Germany), the
Istituto Nazionale di Fisica Nucleare (Italy),
the Research Council of Norway, the
Ministry of Science and Technology of the Russian Federation, and the
Particle Physics and Astronomy Research Council (United Kingdom). 
Individuals have received support from the Swiss 
National Science Foundation, the A. P. Sloan Foundation, 
the Research Corporation,
and the Alexander von Humboldt Foundation.

\end{document}